\documentclass[runningheads]{llncs}
\usepackage{graphicx}
\usepackage{textcomp}
\usepackage{comment}
\usepackage{todonotes}
\usepackage{color}
\usepackage{array}
\newcolumntype{L}[1]{>{\raggedright\let\newline\\\arraybackslash\hspace{0pt}}m{#1}}
\newcolumntype{C}[1]{>{\centering\let\newline\\\arraybackslash\hspace{0pt}}m{#1}}
\newcolumntype{R}[1]{>{\raggedleft\let\newline\\\arraybackslash\hspace{0pt}}m{#1}}
\usepackage{subcaption}
\usepackage{float}

\begin{document}

\title{Quality-aware Cine Cardiac MRI Reconstruction and Analysis from Undersampled k-space Data} 

\author{**\\
**}
\institute{**\\}
\newcommand*\samethanks[1][\value{footnote}]{\footnotemark[#1]}
\newcommand{\rowstyle}[1]{\gdef\currentrowstyle{#1}%
  #1\ignorespaces
}

\author{In\^{e}s Machado* \inst{1} \and 
Esther Puyol-Ant\'on \inst{1}  \and
Kerstin Hammernik \inst{2,3} \and 
Gast\~{a}o Cruz \inst{1} \and 
Devran Ugurlu \inst{1} \and 
Bram Ruijsink \inst{1,4} \and 
Miguel Castelo-Branco \inst{5} \and 
Alistair Young \inst{1} \and 
Claudia Prieto \inst{1} \and 
Julia A. Schnabel \inst{1,2,6} \and 
Andrew P. King \inst{1}} 
\authorrunning{In\^{e}s Machado et al.}   

\institute{School of Biomedical Engineering \& Imaging Sciences, King's College London, UK \and Technical University of Munich, Germany \and Biomedical Image Analysis Group, Imperial College London, UK \and Department of Adult and Paediatric Cardiology, Guy’s and St Thomas’ NHS Foundation Trust, London, UK \and Coimbra Institute for Biomedical Imaging and Translational Research, University of Coimbra, Portugal \and Helmholtz Center Munich, Germany}

\maketitle              
\begin{abstract} 
Cine cardiac MRI is routinely acquired for the assessment of cardiac health, but the imaging process is slow and typically requires several breath-holds to acquire sufficient k-space profiles to ensure good image quality. Several undersampling-based reconstruction techniques have been proposed during the last decades to speed up cine cardiac MRI acquisition. However, the undersampling factor is commonly fixed to conservative values before acquisition to ensure diagnostic image quality, potentially leading to unnecessarily long scan times. In this paper, we propose an end-to-end quality-aware cine short-axis cardiac MRI framework that combines image acquisition and reconstruction with downstream tasks such as segmentation, volume curve analysis and estimation of cardiac functional parameters. The goal is to reduce scan time by acquiring only a fraction of k-space data to enable the reconstruction of images that can pass quality control checks and produce reliable estimates of cardiac functional parameters. The framework consists of a deep learning model for the reconstruction of 2D+t cardiac cine MRI images from undersampled data, an image quality-control step to detect good quality reconstructions, followed by a deep learning model for bi-ventricular segmentation, a quality-control step to detect good quality segmentations and automated  calculation  of  cardiac functional parameters. To demonstrate the feasibility of the proposed approach, we perform simulations using a cohort of selected participants from the UK Biobank (n=270), 200 healthy subjects and 70 patients with cardiomyopathies. Our results show that we can produce quality-controlled images in a scan time reduced from 12 to 4 seconds per slice, enabling reliable estimates of cardiac functional parameters such as ejection fraction within 5\% mean absolute error. \\ 

\keywordname{ Cardiac MRI; deep learning reconstruction; accelerated MRI; image segmentation; quality assessment.} \\
\end{abstract} 

\section{Introduction} 
\hspace{\parindent}

\noindent Cardiac MRI is a common imaging modality for assessing cardiovascular diseases, which is the leading cause of death globally. Cine cardiac MRI enables imaging of the heart throughout the cardiac cycle, and is especially useful for quantifying left and right ventricular function by measuring parameters such as ejection fraction (EF), end-diastolic and end-systolic cardiac chamber volumes. However, cine cardiac MRI acquisition is slow and there has been much research interest in accelerating the scan without compromising the high resolution and image quality requirements. One approach that has been used to speed up the scan is to reduce the amount of acquired k-space data. However, reconstructing cine cardiac MRI from undersampled k-space data is a challenging problem, and approaches typically exploit some type of redundancy or assumption in the underlying data to resolve the aliasing caused by sub-Nyquist sampling \cite{bustin2020compressed}. Considerable efforts have been devoted to accelerate the reconstruction of cardiac MRI from undersampled k-space including parallel imaging and compressed sensing \cite{menchon2019reconstruction}. More recently, machine learning reconstruction approaches have been proposed to learn the non-linear optimization process employed in cardiac MRI undersampled reconstruction. In particular, deep learning (DL) techniques have been proposed to learn the reconstruction process from existing data sets in advance, providing a fast and efficient reconstruction that can be applied to all newly acquired data \cite{schlemper2017deep,hammernik2018learning}. In this paper, we assess image quality from reconstructions of undersampled k-space data during acquisition, creating an ‘active’ acquisition process in which only a fraction of k-space data are acquired to enable the reconstruction of an image that can pass automated quality control (QC) checks and produce reliable estimates of cardiac functional parameters. The major contributions of this work are three-fold: 1) to the best of our knowledge, this is the first paper that combines cine cardiac MRI undersampled reconstruction with QC in downstream tasks such as segmentation in a unified framework; 2) our pipeline includes robust pre- and post-analysis QC mechanisms to detect good quality image reconstructions (QC1) and good quality image segmentations (QC2) during active acquisition and 3) we show that quality-controlled cine cardiac MRI images can be reconstructed in a scan time reduced from 12 to 4 seconds per slice, and that image quality is sufficient to allow clinically relevant parameters (EF and left- and right-ventricle chamber volumes) to be automatically estimated within 5\% mean absolute error.

\section{Materials}
\label{sec:materials}

\noindent We evaluate our proposed reconstruction and analysis framework using a cohort of selected healthy (n=200) and cardiomyopathy (n=70) cases from the UK Biobank obtained on a 1.5 Tesla MRI scanner (MAGNETOM Aera, Siemens Healthcare, Erlangen, Germany). The short-axis (SAX) image acquisition typically consists of 10 image slices with a matrix size of 208×187 and a slice thickness of 8 mm, covering both the ventricles from the base to the apex. The in-plane image resolution is 1.8×1.8 mm\textsuperscript{2}, the slice gap is 2 mm, with a repetition time of 2.6 ms and an echo time of 1.10 ms. Each cardiac cycle consists of 50 time frames. More details of the image acquisition protocol can be found in \cite{petersen2016uk}. The  image reconstruction model and the segmentation model were trained using an additional set of 3,975 cine cardiac MRI datasets from the UK Biobank. For these subjects, pixel-wise segmentations of three structures (left-ventricle (LV) blood pool, right-ventricle (RV) blood pool and LV myocardium) for both end-diastolic (ED) frames and end-systolic (ES) frames were manually performed to act as ground truth segmentations \cite{petersen2017reference}.
The segmentations were performed by a group of eight observers and each subject was annotated only once by one observer. Visual QC was performed on a subset of the data to ensure acceptable inter-observer agreement. The segmentation model was evaluated using 600 different subjects from the UK Biobank for intra-domain testing and two other datasets for cross-domain testing: the Automated Cardiac Diagnosis Challenge (ACDC) dataset (100 subjects, 1 site, 2 scanners) and the British Society of Cardiovascular Magnetic Resonance Aortic Stenosis (BSCMR-AS) dataset (599 subjects, 6 sites, 9 scanners).

\section{Methods}

\begin{figure}[h!]
\centering
\includegraphics[width=1.0\textwidth]{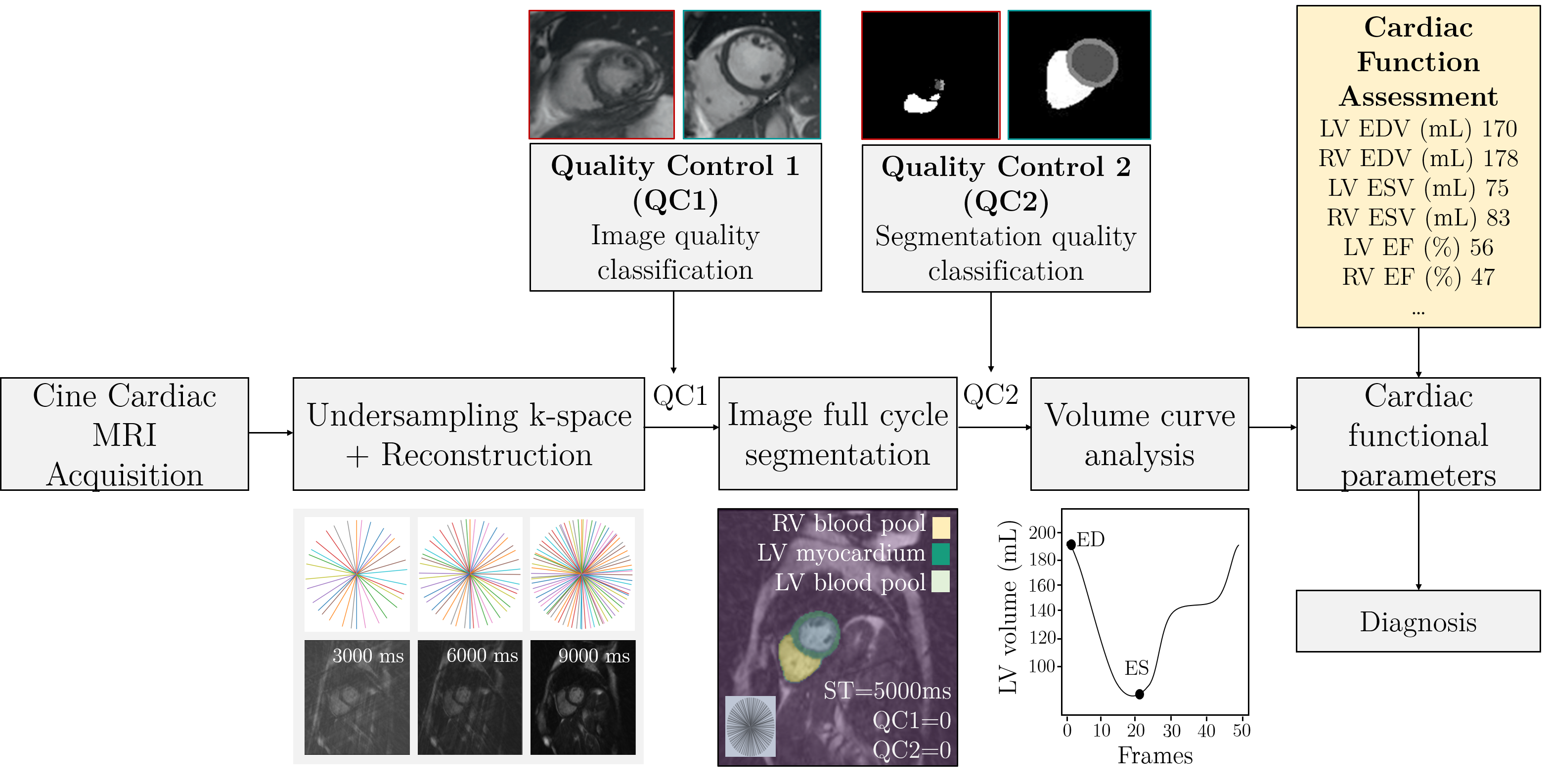}
\caption{Overview of the image analysis pipeline for fully-automated cine cardiac MRI undersampled reconstruction and analysis including comprehensive QC algorithms to detect erroneous output. As k-space profiles are acquired, images are continually reconstructed and passed through QC checks. The simulated acquisition terminates when the reconstructed image
passes all QC checks.}
\label{fig:pipeline}
\end{figure}

The developed image analysis pipeline consists of a DL model for accelerated reconstruction of SAX cine cardiac MRI acquisitions, a DL model for automatic segmentation of the LV blood pool, RV blood pool and LV myocardium, automated calculation of cardiac functional parameters such as EF and LV and RV chamber volumes, and two QC steps: a pre-analysis image QC step during the undersampling and reconstruction process (QC1) and a segmentation QC step to detect good quality segmentations (QC2). For an illustration of the pipeline see Figure \ref{fig:pipeline}. \\

\noindent\textbf{Undersampling and reconstruction:} We simulated an active acquisition process by first creating k-space data from all slices of cine SAX cardiac MRI images. We utilised a similar strategy to \cite{haldar2013low} to generate synthetic phase and a radial golden angle sampling pattern to simulate undersampled k-space data containing increasing numbers of profiles corresponding to scan times between 1 to 30 seconds, in steps of 1 second. These were then reconstructed using two reconstruction algorithms for comparison: the non-uniform Fast Fourier Transform (nuFFT) and the Deep Cascade of Convolutional Neural Networks (DCCNN) \cite{schlemper2017deep,hammernik2018learning} which features alternating data consistency layers and regularisation layers within an unrolled end-to-end framework. Undersampled k-space data, along with the sampling trajectory and density compensation function, are provided as input to this unrolled model for DL reconstruction, and high-quality MRI images are obtained as an output in an end-to-end fashion. The regularisation layers of this network were implemented as a 5-layer CNN according to \cite{schlemper2017deep}, and the data consistency layers follow a gradient descent scheme according to \cite{hammernik2018learning}. \\

\noindent\textbf{Image quality control (QC1):} QC1 was framed as a binary classification problem and addressed using a ResNet classification network \cite{he2016deep}. Binary image quality labels (analyzable/non-analyzable) from 225 images of different levels of undersampling from UK Biobank subjects were generated by visual inspection and validated by an expert cardiologist. 80\% were used for training and validation of the classification network and 20\% were used for testing. The ResNet network was trained for 200 epochs with a binary cross entropy loss function. During training, data augmentation was performed on-the-fly including rotation, shifts and image intensity transformations. The probability of augmentation for each of the parameters was 50\%. The training/testing images for QC1 were randomly selected from the UK Biobank dataset and were not used for training or evaluating the reconstruction/analysis framework.
The training/testing dataset for QC1 consisted of 50\% healthy subjects and 50\% patients with cardiomyopathies. \\

\noindent\textbf{Image full cycle segmentation:} We used a pre-trained U-net based architecture \cite{chen2020improving}
for automatic segmentation of the LV blood pool, LV myocardium and RV blood pool from all SAX slices and all frames through the cardiac cycle. The UK Biobank dataset was split into three subsets, containing 3975, 300 and 600 subjects for training, validation and testing respectively. All images were resampled to 1.25 × 1.25 mm. The training dataset was augmented in order to cover a wide range of geometrical variations in terms of the heart pose and size. All images were cropped to the same size of 256×256 before being fed into the network. \\

\noindent\textbf{Segmentation quality control (QC2):} The segmentation process is followed by a segmentation QC step (QC2) based on Reverse Classification Accuracy (RCA) \cite{robinson2019automated}. First, image registration is performed between the test image and a set of 20 pre-selected template images with known segmentations. Next, the transformed test image segmentation is compared to those of the atlas segmentations, and a high similarity is assumed to indicate a good quality test image segmentation. The segmentation quality metrics used were the Dice Similarity Coefficient (DSC), Mean Surface Distance (MSD), Root-Mean-Square Surface Distance (RMSD) and Hausdorff Distance (HD). Finally, a SVM binary classifier was trained using the quality metrics independently for LV and RV, and ED and ES frames to discriminate between poor and good quality segmentations. \\

\noindent\textbf{Clinical functional parameters:} The volumes were calculated by multiplying the number of voxels by the voxel volume for each of the LV/RV classes. The maximum volume over the cardiac cycle was used for (LV/RV)EDV and the minimum for (LV/RV)ESV. EF (for both LV and RV) was calculated as (EDV-ESV)/EDV. \\

\section{Results}

We validated our method in two ways. First, we evaluated the ability of the full pipeline to detect good quality image reconstructions (QC1) and good quality image segmentations (QC2) during simulated active acquisition of 270 cases (200 healthy subjects and 70 patients with cardiomyopathies) randomly selected from the UK Biobank cohort (Validation 1). Second, we compared the estimates of cardiac functional parameters obtained via our pipeline to those obtained from the fully-sampled data (Validation 2). \\

\noindent\textbf{Validation 1.} Image quality was evaluated with the Mean Absolute Error (MAE), Peak Signal to Noise Ratio (PSNR) and Structural Similarity Index (SSIM), calculated between the fully-sampled image and the undersampled image that passed all QC checks with the lowest scan time. Segmentation quality was quantified using Dice coefficients between the segmentations from the fully-sampled image (obtained using the U-net) and the segmentations derived from our pipeline. Image and segmentation quality results are shown in Table \ref{table:results}.\\

\begin{figure}[t] 
  \centering
  \includegraphics[width=1.0\textwidth]{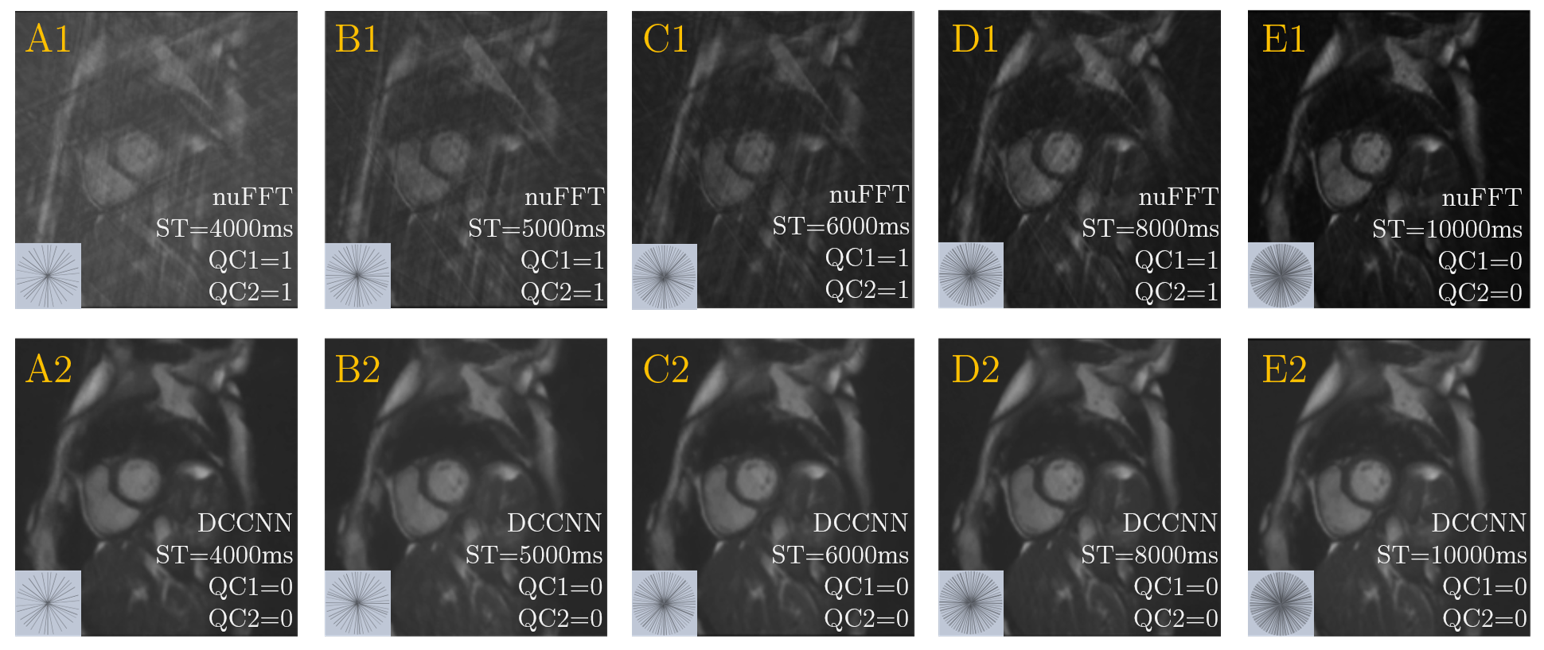}
  \caption{Illustration of image reconstructions and undersampling trajectories as a function of the scan time using nuFFT and DCCNN. For this subject, the two QC steps were passed at a scan time of 10 seconds and 4 seconds with the nuFFT and DCCNN reconstruction models respectively. (QC=0 means that the QC check was passed.)}
    \label{fig:reconstructions}
\end{figure}

\begin{figure}[H] 
  \centering
  \includegraphics[width=1.0\textwidth]{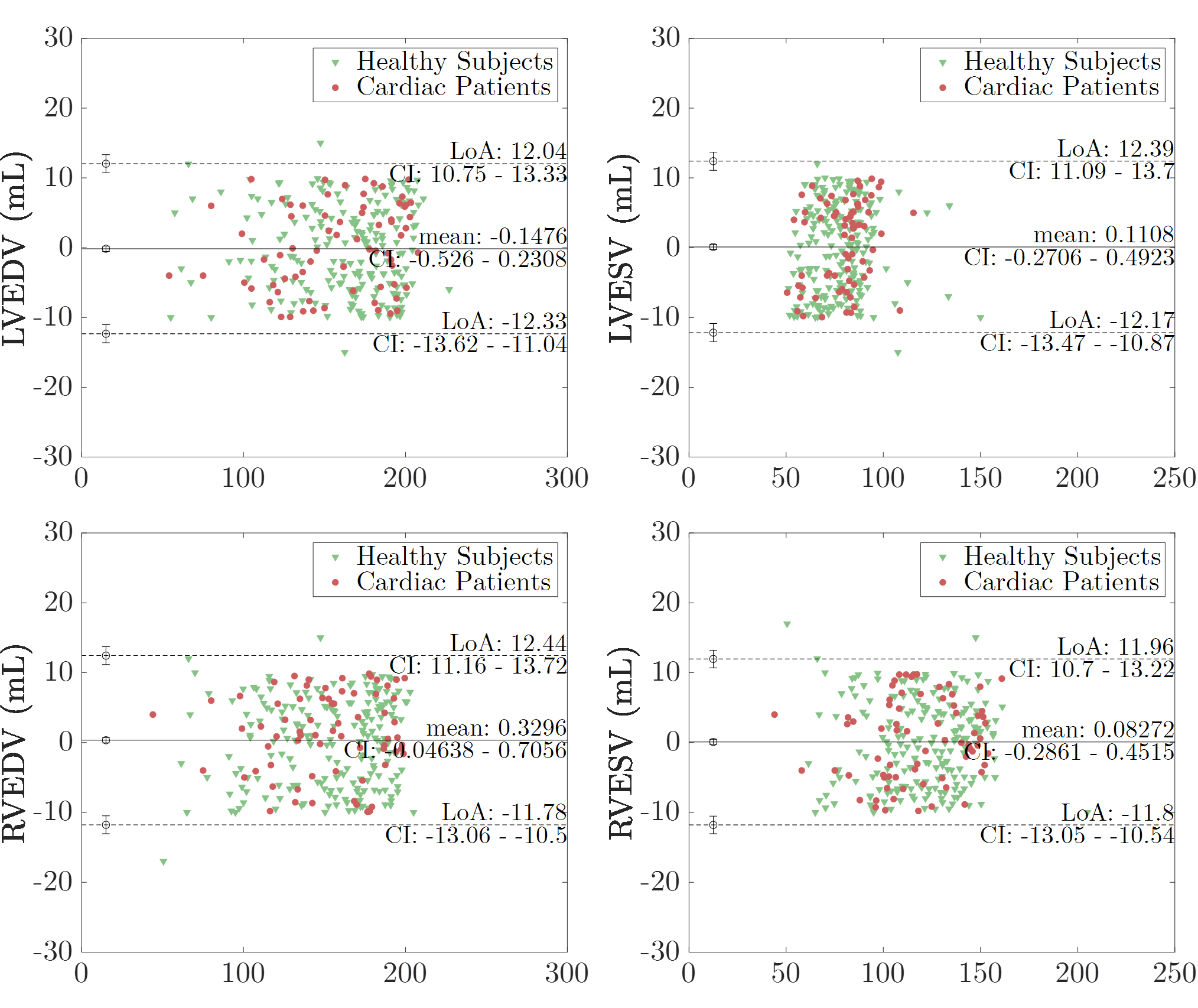}
  \caption{Bland-Altman plots for cardiac volumes between U-net segmentations from fully sampled reconstructions and our DCCNN reconstructions with smallest scan time that passed all QC checks. The black solid line represents the mean bias and the black dotted lines the limits of agreement. The limits of agreement are defined as the mean difference ± 1.96 SD of differences.}
  \label{fig:plots}
\end{figure}

\noindent\textbf{Validation 2.} The performance of the method was also evaluated using clinically relevant measures: LVEDV, LVESV, LVEF, RVEDV, RVESV and RVEF. Our DCCNN-based approach resulted in a closer match to the fully-sampled data measures than the nuFFT method and also resulted in a lower scan time as shown in Table \ref{table:results}. There was a good correlation between estimations obtained from fully-sampled data and via our pipeline (Pearson's correlations: LVEDV: r = 0.98; LVESV: r = 0.97; LVEF: r = 0.98; RVEDV: r = 0.98; RVESV: r = 0.95 and RVEF: r = 0.96). Figure \ref{fig:reconstructions} illustrates image reconstructions and undersampling trajectories as a function of the scan time using nuFFT and DCCNN. A Bland-Altman analysis between the volumes estimated from fully-sampled reconstructions and using our DCCNN-based pipeline is shown in Figure \ref{fig:plots}. 

\begin{table} [t] 
\centering
\caption{Top: Sensitivity, specificity, and average balanced accuracy for QC1. Middle: Image quality of 270 subjects from the UK Biobank after passing QC1. Bottom: Dice scores between the segmentations from the fully-sampled images and the segmentations derived from our pipeline. The mean and standard deviation of scan times at which the reconstructed images passed all QC checks are also reported.}
\renewcommand{\arraystretch}{0.3}
\begin{tabular}{L{6cm}L{2.8cm}L{2.8cm}}

\\ \hline 
&  & \\
& {\bf nuFFT} & {\bf DCCNN} \\ 
&  &  \\
\\ \hline 
&  & \\
{\bf QC1 Training model} &  & \\ 
&  &  \\
&  &  \\
Average balanced accuracy &  0.93 & 0.95 \\
&  &  \\
Sensitivity & 0.86 & 0.87 \\ 
&  &  \\
Specificity & 0.99 & 0.99 \\
&  &  \\
\\ \hline 
&  & \\
{\bf QC1 Image quality metrics} &  & \\ 
&  &  \\ 
&  &  \\
MAE &  $0.03 \pm 0.02$ & $0.04 \pm 0.02$ \\
&  &  \\
PSNR & $29.82 \pm 0.08$ & $32.14 \pm 0.06$ \\
&  &  \\
SSIM & $0.87 \pm 0.04$ & $0.90 \pm 0.03$ \\
&  &  \\ 
&  & \\ \hline 
&  & \\
{\bf QC2 Dice scores} &  & \\ 
&  &  \\ 
&  &  \\
LV blood pool &  $0.91 \pm 0.06$ & $0.93 \pm 0.04$  \\ 
&  &  \\
LV myocardium & $0.91 \pm 0.06$ & $0.94 \pm 0.05$ \\ 
&  &  \\
RV blood pool & $0.89 \pm 0.04$  & $0.90 \pm 0.06$  \\
&  &  \\
\\ \hline
&  &  \\
{\bf Scan Time (s)} & $11.82 \pm 3.29$ & $3.72 \pm 0.54$ \\
&  &  \\
\hline 
\end{tabular}
\label{table:results}
\end{table}

\section{Discussion}
\label{sec:discussion}
This work has demonstrated the feasibility of a DL-based framework for automated quality-controlled “active” acquisition of undersampled cine cardiac MRI data without a previously defined undersampling factor. The proposed pipeline results in a reduced scan time for 2D cardiac cine MRI, which takes $\sim$12 s in our clinical protocol (spatial resolution = 1.8 x 1.8 x 8.0 mm$^3$, temporal resolution = 31.56 ms and undersampling factor = 2). Our results show that by using a
DCCNN for cine cardiac MRI reconstruction, we can pass QC checks after approximately 4 seconds of simulated acquisition (i.e. an undersampling factor of 4.5). One limitation of our approach is that in the current version of our pipeline, QC2 takes $\sim$1min. Therefore, this would not allow immediate quality feedback and true ‘active’ acquisition. Future investigations will focus on developing a

\noindent real-time approach for segmentation quality control. For example, in \cite{galati2021efficient}, a convolutional autoencoder was trained to quantify segmentation quality without a ground truth at inference time. In \cite{robinson2018real}, a 3D convolutional neural network was trained in order to predict the DSC values of 3D segmentations. We will investigate these and similar approaches to implement a segmentation quality control step that is efficient enough to facilitate real-time implementation on the MRI scanner. We believe that an approach such as the one we have proposed could have great clinical utility, reducing redundancies in the cardiac MRI acquisition process whilst still providing diagnostic quality images and robust estimates of functional parameters.


\section*{Acknowledgements}
This work was funded by the Engineering and Physical Sciences Research Council (EPSRC) programme grant ‘SmartHeart’ (EP/P001009/1) and supported by the Wellcome/EPSRC Centre for Medical Engineering [WT 203148/Z/16/Z]. The research was supported by the National Institute for Health Research (NIHR) Biomedical Research Centre based at Guy's and St Thomas' NHS Foundation Trust and King's College London. The views expressed are those of the author(s) and not necessarily those of the NHS, the NIHR or the Department of Health.
This work was also supported by Health Data Research UK, an initiative funded by UK Research and Innovation, Department of Health and Social Care (England) and the devolved
administrations, and leading medical research charities. This research has been conducted using the UK Biobank Resource under Application Number 17806.


\bibliographystyle{splncs03unsrt}
\bibliography{refs}

\end{document}